\documentclass[reprint,showpacs,amsmath,amssymb,aps,usenatbib]{revtex4-1}
\usepackage{graphicx}
\usepackage{dcolumn}
\usepackage{bm}
\usepackage{subfigure}
\usepackage{longtable}
\usepackage{setspace}
\usepackage[bookmarksnumbered,bookmarksopen,colorlinks,citecolor=blue,linkcolor=blue]{hyperref}
\usepackage{natbib}
\usepackage{amsmath}
\usepackage{rotating}
\usepackage{colortbl}
\usepackage{color}
\definecolor{mygray}{gray}{0.8}

\begin{document}
\title{New quantification of symmetry energy from neutron skin thicknesses of $^{48}$Ca and $^{208}$Pb}

\author{Rong An}
\affiliation{School of Physics, Ningxia University, Yinchuan 750021, China}
\affiliation{Key Laboratory of Beam Technology of Ministry of Education, College of Nuclear Science and Technology, Beijing Normal University, Beijing 100875, China}
\author{Shuai Sun}
\affiliation{Key Laboratory of Beam Technology of Ministry of Education, College of Nuclear Science and Technology, Beijing Normal University, Beijing 100875, China}
\author{Li-Gang Cao}
\email[Corresponding author: ]{caolg@bnu.edu.cn}
\affiliation{Key Laboratory of Beam Technology of Ministry of Education, College of Nuclear Science and Technology, Beijing Normal University, Beijing 100875, China}
\affiliation{Key Laboratory of Beam Technology of Ministry of Education, Institute of Radiation Technology, Beijing Academy of Science and Technology, Beijing 100875, China}
\author{Feng-Shou Zhang}
\email[Corresponding author: ]{fszhang@bnu.edu.cn} 
\affiliation{Key Laboratory of Beam Technology of Ministry of Education, College of Nuclear Science and Technology, Beijing Normal University, Beijing 100875, China}
\affiliation{Key Laboratory of Beam Technology of Ministry of Education, Institute of Radiation Technology, Beijing Academy of Science and Technology, Beijing 100875, China}
\affiliation{Center of Theoretical Nuclear Physics, National Laboratory of Heavy Ion Accelerator of Lanzhou, Lanzhou 730000, China}

\begin{abstract}
Precise knowledge of the nuclear symmetry energy can be tentatively calibrated through multimessenger constraints.
The neutron skin thickness of a heavy nucleus is one of the most sensitive indicators for probing the isovector components of effective interactions in asymmetric nuclear matter.
Recent studies have suggested that the experimental data from the CREX and PREX2 Collaborations are not mutually compatible within existing nuclear models.
In this study, we review the quantification of the slope parameter of symmetry energy $L$ from the neutron skin thicknesses of $^{48}$Ca and $^{208}$Pb.
Skyrme energy density functionals classified by various isoscalar incompressibility coefficients $K$ are employed to evaluate the bulk properties of finite nuclei.
The calculated results suggest that the slope parameter $L$ deduced from $^{208}$Pb is sensitive to the compression modulus of symmetric nuclear matter, but not that from $^{48}$Ca.
The effective parameter sets classified by $K=220$ MeV can provide an almost overlaping range of $L$ from $^{48}$Ca and $^{208}$Pb.
\end{abstract}

\maketitle

\section{Introduction}\label{section1}
Nuclear symmetry energy (NSE), which characterizes the energy cost of converting isospin symmetric nuclear matter (SNM) into pure neutron matter (PNM), plays a vital role in determining the properties of finite nuclei and neutron stars~\cite{PhysRevLett.86.5647,Li2019,Steiner:2004fi,LATTIMER2007109,Xiachengjun}. The density dependence of the NSE, that is, $E_{\mathrm{sym}}(\rho)$, can be expanded around the saturation density $\rho_{0}$ ($\simeq0.16$~fm$^{-3}$).
The slope parameter $L$ dominates the behavior of the equation of state (EoS) for asymmetric nuclear matter in the vicinity of $\rho_{0}$.
Precise knowledge about the density dependence of the NSE is difficult to obtain owing to the uncertainties arising from the varying model-dependent slope parameter $L$.
Fortunately, the characteristic behaviors of the NSE can be extracted indirectly from both extensive terrestrial nuclear experiments and observed astrophysical events~\cite{Li:2008gp,Li:2014oda,Hu:2020ujf,PhysRevC.90.064317,PhysRevC.101.034303,Liu_2018,PhysRevC.108.L021303}.

Nuclear symmetry energy has been extensively used to encode the implications of the degree of isospin-asymmetry in finite nuclei.
This is especially useful in the formation of the neutron skin thickness (NST) or neutron halo structure~\cite{PhysRevLett.77.3963,PhysRevC.82.011301,PhysRevC.102.044313}.
The quantity of NST $\Delta{R_{\mathrm{np}}}=\sqrt{\langle{r_{n}^{2}}\rangle}-\sqrt{\langle{r_{p}^{2}}\rangle}$, which is defined as the difference between the root-mean-square (rms) radii of the neutrons and protons in a heavy nucleus and is strongly correlated to the slope parameter of the NSE, $L$~\cite{Brown20005296,PhysRevC.64.027302,PhysRevC.69.024318,PhysRevC.72.064309,PhysRevLett.106.252501,PhysRevC.80.024316,
PhysRevC.84.034316,PhysRevLett.102.122502,PhysRevC.81.051303,PhysRevLett.109.262501,PhysRevC.87.051306,PhysRevC.87.034327,
PhysRevC.93.051303,ZHANG2013234,PhysRevC.102.044316,particles6010003,PhysRevC.91.034315,PhysRevC.90.064310,chen2006,PhysRevC.97.034318}.
Therefore, the NST of a heavy nucleus was undertaken to provide a constraint on the EoS of neutron-rich matter around $\rho_{0}$.

The neutron radius of $^{208}$Pb has been determined in a laboratory by measuring the parity violating asymmetry $A_{\mathrm{PV}}$ in polarized elastic electron scattering experiments such as PREX2~\cite{PhysRevLett.126.172502}.
These efforts provided the latest value of NST with significantly improved precision: $\Delta{R}_{\mathrm{np}}^{208}=0.212\sim0.354$ fm.
Moreover, a precise measurement of the NST for $^{48}$Ca was updated by the CREX group:  $\Delta{R}_{\mathrm{np}}^{48}=0.071\sim0.171$ fm~\cite{PhysRevLett.129.042501}.
The reported NST of $^{48}$Ca is relatively thin compared to the measurement obtained by the high-resolution electric polarizability experiment ($\alpha_{D}$) in the RCNP collaboration ($\Delta{R}_{\mathrm{np}}^{48}=0.14\sim0.20$ fm)~\cite{PhysRevLett.118.252501}.
In contrast, the NST of $\Delta{R}_{\mathrm{np}}^{208}$ obtained by the PREX2 Collaboration is larger than that measured by RCNP ($\Delta{R}_{\mathrm{np}}^{208}=0.135\sim0.181$ fm)~\cite{PhysRevLett.107.062502}.
In Ref.~\cite{PhysRevLett.127.192701}, the neutron skin thickness of $^{208}$Pb obtained by constraining the astrophysical observables favors smaller value; for example, $\Delta{R}_{\mathrm{np}}^{208}=0.17\pm0.04$ fm.
Likewise, the optimized new functionals obtained by calibrating the $A_{\mathrm{PV}}$ and $\alpha_{D}$ values of $^{208}$Pb predict an NST of $\Delta{R}_{\mathrm{np}}^{208}=0.19\pm0.02$ fm and the symmetry-energy slope $L=54\pm8$ MeV~\cite{PhysRevLett.127.232501}.
Recently theoretical studies have suggested that neutron star masses and radii are more sensitive to the NST of $^{208}$Pb than its dipole polarizability $\alpha_{D}$~\cite{PhysRevC.107.035802}.
These results challenge our understanding of nuclear force and energy density functionals (EDFs).

In Ref.~\cite{TAGAMI2022106037}, 207 EoSs were employed to explore the systematic correlations between $\Delta{R}_{\mathrm{np}}^{48}$ and $L$(CREX) and between $\Delta{R}_{\mathrm{np}}^{208}$ and $L$(PREX2).
The slope parameter of the NSE obtained by fitting $\Delta{R}_{\mathrm{np}}^{48}$ covers the interval range $L(\mathrm{CREX})=0\sim50$ MeV;
however, the calibrated correlation between the slope parameter $L$ and $\Delta{R}_{\mathrm{np}}^{208}$ yields $L$(PREX2)$=76\sim165$ MeV.
As mentioned in the literature, there is no overlap between $L$(CREX) and $L$(PREX2) at the one-$\sigma$ level.
A combined analysis was also performed using a recent experimental determination of the parity violating asymmetry in $^{48}$Ca and $^{208}$Pb~\cite{PhysRevLett.129.232501}.
The study demonstrated that the existing nuclear EDFs cannot simultaneously offer an accurate description of the skins of $^{48}$Ca and $^{208}$Pb.
The same scenario can also be encountered in Bayesian analysis, where the predicted $\Delta{R}_{\mathrm{np}}^{48}$ is closed to the CREX result, but considerably underestimates the result of $\Delta{R}_{\mathrm{np}}^{208}$ with respect to the PREX2 measurement~\cite{Zhang:2022bni}.
Considering the isoscalar-isovector couplings in relativistic EDFs, the constraints from various high-density data cannot reconcile the recent results from PREX2 and CREX Collaboration measurements~\cite{Miyatsu:2023lki}.
These investigations indicate that it is difficult to provide consistent constraints for the isovector components of the EoSs using existing nuclear EDFs, and further theoretical and experimental studies are urgently required~\cite{YUKSEL2023137622}.

To reduce the discrepancies between the different measurements and observations, an extra term controlling the dominant gradient correction to the local functional in the isoscalar sector has been used to weaken the correlations between the properties of finite nuclei and the nuclear EoS~\cite{PhysRevC.107.015801}.
As demonstrated in Ref.~\cite{PhysRevC.96.065805}, the influence of the isoscalar sector is nonnegligible in the analysis.
Nuclear matter properties expressed in terms of their isoscalar and isovector counterparts are correlated~\cite{PhysRevC.73.044320}.
As noticed above, existing discussions focus on the isovector components in the EDFs model.
Characteristic isoscalar quantities, such as the incompressibility of symmetric nuclear matter, are less considered when determining the slope parameter $L$~\cite{TAGAMI2022106037}.
The nuclear incompressibility can be deduced from measurements of the isoscalar giant monopole resonance (ISGMR) in medium-heavy nuclei~\cite{PhysRevC.70.024307,PhysRevC.86.054313} and multi-fragmentations of heavy ion collisions~\cite{PhysRevC.70.041604}.
The NSE obtained through the effective Skyrme-EDF is related to the isoscalar and isovector effective masses, which are also indirectly related to the incompressibility of symmetric nuclear matter~\cite{PhysRevC.73.014313}.
Although correlations between the incompressibility coefficients and isovector parameters are generally weaker than correlations between the slope parameter $L$ and NSE~\cite{PhysRevC.104.054324}, quantification uncertainty due to nuclear matter incompressibility is inevitable in the evaluation.
Therefore, the influence of the isoscalar nuclear matter properties is essential for evaluating slope parameter $L$.

The remainder of this paper is organized as follows. In Sec.~\ref{sec1}, we briefly describe our theoretical model. In Sec.~\ref{sec2}, we present the results and discussion. A short summary and outlook are provided in Sec.~\ref{sec3}.

\section{Theoretical framework}\label{sec1}
The sophisticated Skyrme-EDF, expressed as an effective zero-range force between nucleons with density- and momentum-dependent terms, has been succeeded in describing various physical phenomena~\cite{RevModPhys.75.121,PhysRevC.95.014316,PhysRevC.102.054312,PhysRevC.90.024317,PhysRevC.102.014312,WU2022136886,
Caotens11,Wenpw14,PhysRevC.87.064311,Caoquench}.
In this study, Skyrme-like effective interactions were calculated as follows~\cite{CHABANAT1997710,CHABANAT1998231}:
\begin{eqnarray}
V(\mathbf{r}_{1},\mathbf{r}_{2})&=&t_{0}(1+x_{0}\mathbf{P}_{\sigma})\delta(\mathbf{r})\nonumber\\
&&+\frac{1}{2}t_{1}(1+x_{1}\mathbf{P}_{\sigma})\left[\mathbf{P}'^{2}\delta(\mathbf{r})+\delta(\mathbf{r})\mathbf{P}^{2}\right]\nonumber\\
&&+t_{2}(1+x_{2}\mathbf{P}_{\sigma})\mathbf{P}'\cdot\delta(\mathbf{r})\mathbf{P}\nonumber\\
&&+\frac{1}{6}t_{3}(1+x_{3}\mathbf{P}_{\sigma})[\rho(\mathbf{R})]^{\alpha}\delta(\mathbf{r})\nonumber\\
&&+\mathrm{i}W_{0}\mathbf{\sigma}\cdot\left[\mathbf{P}'\times\delta(\mathbf{r})\mathbf{P}\right],
\end{eqnarray}
where $\mathbf{r}=\mathbf{r}_{1}-\mathbf{r}_{2}$ and $\mathbf{R}=(\mathbf{r}_{1}+\mathbf{r}_{2})/2$ are related to the positions of two nucleons $\mathbf{r}_{1}$ and $\mathbf{r}_{2}$, $\mathbf{P}=(\nabla_{1}-\nabla_{2})/2\mathrm{i}$ is the relative momentum operator and $\mathbf{P'}$ is its complex conjugate acting on the left, and $\mathbf{P_{\sigma}}=(1+\vec{\sigma}_{1}\cdot\vec{\sigma}_{2})/2$ is the spin exchange operator that controls the relative strength of the $S=0$ and $S=1$ channels for a given term in the two-body
interactions, where $\vec{\sigma}_{1(2)}$ are the Pauli matrices.
The final term denotes the spin-orbit force, where $\sigma=\vec{\sigma}_{1}+\vec{\sigma}_{2}$.
Quantities $\alpha$, $t_{i}$ and $x_{i}$ ($i=0$-3) represent the effective interaction parameters of the Skyrme forces.

Generally, effective interaction parameter sets are calibrated by matching the properties of finite nuclei and nuclear matter at the saturation density.
Notably,the Skyrme-EDF can provide an analytical expression of all variables characterizing infinite nuclear matter (see~\cite{CHABANAT1997710,CHABANAT1998231,PhysRevC.85.035201,PhysRevC.82.024321} for details).
The neutron skin of a heavy nucleus is regarded as the feasible indicator for probing the isovector interactions in the EoS of asymmetric nuclear matter.
Thus, the neutron and proton density distributions can be self-consistently  calculated using Skyrme DEFs with various parameter sets.
To clarify this, we further inspected the correlations between the slope parameter $L$ and the NSTs of $^{48}$Ca and $^{208}$Pb. The bulk properties were calculated using the standard Skyrme-type EDFs~\cite{PhysRevC.70.024307}.
The corresponding effective interactions were in accord with the calculated nuclear matter properties, such as binding energy per nucleon $E=\mathcal{E}/{\rho}$, symmetry energy $E_{\mathrm{sym}}(\rho)=\frac{1}{8}\partial^{2}(\mathcal{E}/\rho)/\partial\rho^{2}|_{\rho=\rho_{0}}$, slope parameter $L=3\rho_{0}\partial{E_{\mathrm{sym}}(\rho)}/{{\partial\rho}}|_{\rho=\rho_{0}}$, and the incompressibility coefficient $K=9\rho_{0}^{2}\partial^{2}(\mathcal{E}/\rho)/{\partial\rho^{2}}|_{\rho=\rho_{0}}$.
The value of the isoscalar incompressibility $K$ from experimental data on giant monopole resonances covers a range of $230\pm10$ MeV~\cite{PhysRevLett.82.691,PhysRevC.69.051301}.
In addition, the incompressibility of symmetric nuclear matter deduced from $\alpha$-decay properties is $K=241.28$ MeV~\cite{PhysRevC.74.034302}.

\begin{table}[htbp!]
\centering
\caption{Saturation properties with the different Skyrme parameter sets, such as symmetry energy $E_{\mathrm{sym}}$ (MeV), the slope parameter $L$ (MeV) and the nuclear matter incompressibility coefficient $K$ (MeV) at saturation density $\rho_{0}$ (fm$^{-3}$), are shown definitely~\cite{PhysRevC.70.024307,PhysRevC.85.035201}.}\label{tab0}
\begin{tabular}{cccc}
\hline
\hline
 ~~~$K$~(MeV)~~~ & ~~~~Sets~~~~ & ~~~$E_{_{\mathrm{sym}}}$~(MeV)~~~& ~~~~$L$~(MeV)~~~~ \\
\hline
&s2028  & 28 & ~~5.21  \\
&s2030  & 30 & ~12.20  \\
&s2032  & 32 & ~33.31  \\
$K$ =220 &s2034  & 34 & ~40.37  \\
&s2036  & 36 & ~58.82  \\
&s2038  & 38 & ~72.59  \\
&s2040  & 40 & ~83.22  \\
\hline
&s3028   & 28 & $-$11.23~ \\
&s3030  & 30 & ~22.87  \\
&s3032  & 32 & ~36.22  \\
$K$ =230  &s3034  & 34 & ~56.14  \\
&s3036  & 36 & ~71.54  \\
&s3038  & 38 & ~87.62  \\
&s3040  & 40 & 106.09  \\
\hline
&s4028  &28  &~~3.98  \\
&s4030  &30  &~34.07  \\
&s4032  &32  &~34.43  \\
$K$ =240 &s4034  &34  &~62.59  \\
&s4036  &36  &~75.67  \\
&s4038  &38  &~98.65  \\
&s4040  &40  &108.17  \\
\hline\hline
\end{tabular}
\end{table}
The nuclear breathing model exhibits a moderate correlation with the slope of the NSE and a strong dependence on the isoscalar incompressibility coefficient $K$ of the symmetric nuclear matter~\cite{Chen_2012}.
The incompressibility of nuclear matter helps us understand the properties of neutron stars~\cite{PhysRevC.94.052801,PhysRevC.104.055804}.
Thus, it is essential to inspect the influence of isoscalar components on the slope parameter of the symmetry energy.
To facilitate a quantitative discussion, a series of effective interaction sets classified by various nuclear incompressibility coefficients ($K=220$ MeV, 230 MeV, and 240 MeV)were employed, as shown in Table~\ref{tab0}.
Generally, analytical expressions at the saturation density $\rho_{0}$ have specific forms~\cite{PhysRevC.85.035201}.
Using these expressions, the density dependence of the symmetry energy can be expanded as a function of neutron excess.
Under the corresponding $K$, the slope parameter $L$ and symmetry energy $E_{\mathrm{sym}}$ at the saturation density $\rho_{0}$ also cover a large range.

\section{Results and Discussions}\label{sec2}
\begin{figure}[htbp]
  \centering
    \includegraphics[width=1.1\linewidth]{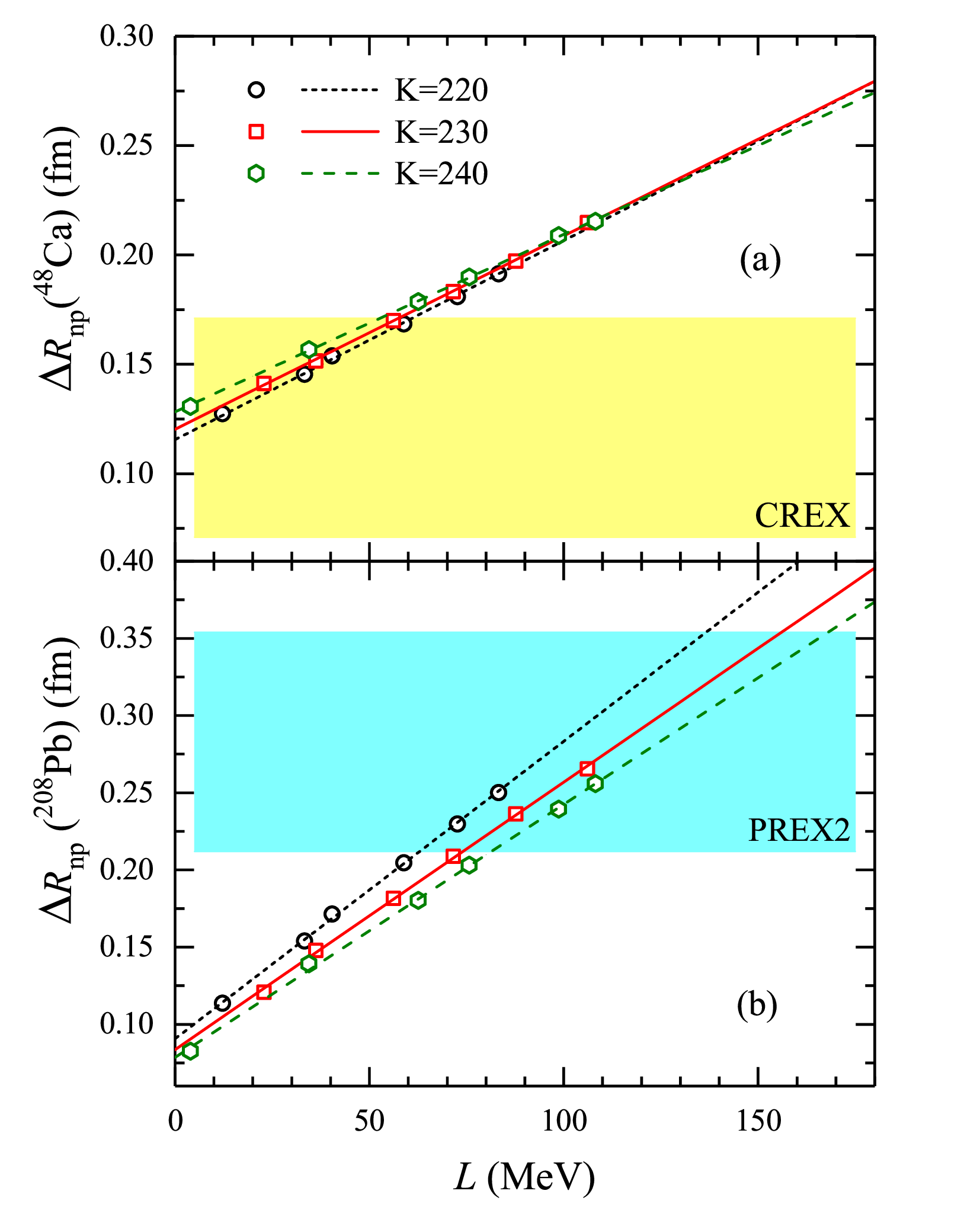}
    \caption{(Color online) Neutron skin thickness of $^{48}$Ca and $^{208}$Pb as a function of slope parameter $L$ at saturation density $\rho_{0}$. Experimental constraints are indicated by horizontal light-yellow (a) and blue (b) bands. The open markers represent Skyrme-EDFs calculations classified by various incompressibility coefficients. The corresponding lines indicate theoretical linear fits.}\label{fig1}
\end{figure}

In Fig.~\ref{fig1}, the NSTs of $^{48}$Ca and $^{208}$Pb are determined under various effective interactions.
The chosen parameter sets are classified by different incompressibility coefficients of symmetric nuclear matter, for example, $K=220$ MeV, $230$ MeV and $240$ MeV.
The experimental constraint on the NST is indicated by a colored shadow.
With increasing slope parameter $L$, the NST is increased monotonically, and strong linear correlations between $L$ and the NST of $^{48}$Ca and $^{208}$Pb are observed.
As shown in Fig.~\ref{fig1}~(a), the linear correlations are almost similar, and the gradients for these three lines are in the ranges of $0.0008\sim0.0009$.
Figure~\ref{fig1}~(b) shows the related linear correlations between $\Delta{R_{\mathrm{np}}}$($^{208}$Pb) and the slope parameters $L$ for various nuclear matter incompressibility coefficients.
However, with increasing incompressibility coefficient, the slopes of the fitted lines gradually decrease or a large deviation emerges at a high $L$.
The nuclear matter EoS is conventionally defined as the binding energy per nucleon and can be expressed as Taylor-series expansion in terms of the isospin asymmetry.
As suggested in Refs.~\cite{PhysRevC.69.041301,Chen_2012}, the compression modulus of symmetric nuclear matter is sensitive to the density dependence of the NSE. With increasing neutron star mass, the correlation between $K$ and its slope $L$ increases~\cite{PhysRevC.104.055804}. From this figure, we can see that the isoscalar quantity of the incompressibility coefficient has a significant influence on the determination of the slope parameter $L$ in $^{208}$Pb. However, for $^{48}$Ca this influence can be ignored.

Herein, we assume that the value of $L$ is positive. Linear functions were fitted to the data classified by various nuclear matter incompressibility coefficients using the least-squares method. For $K=220$ MeV, we obtained the $L-\Delta{R_{\mathrm{np}}^{48}}$ relation as
\begin{eqnarray}
\Delta{R_{\mathrm{np}}^{48}}=0.0009L+0.1155>0.1155~~~~\mathrm{fm}.
\end{eqnarray}
For $L-\Delta{R}_{\mathrm{np}}^{208}$, the linear function$K=220$ MeV is expressed as
\begin{eqnarray}
\Delta{R_{\mathrm{np}}^{208}}=0.0019L+0.0914~~~~\mathrm{fm},
\end{eqnarray}
where a high correlation coefficient is located at $R=0.99$.

As suggested in Ref.~\cite{TAGAMI2022106037}, the slope parameter $L$ ($0\sim 50$ MeV) deduced from $\Delta{R}_{\mathrm{np}}^{48}$ cannot overlap the interval range of the slope parameter $L$ ($76\sim 165$ MeV) deduced from $\Delta{R}_{\mathrm{np}}^{208}$.
To facilitate a quantitative comparison of the experiments with these theoretical calculations, the slope parameter $L$ derived from the constraints of the NSTs of $^{48}$Ca and $^{208}$Pb are presented for various nuclear matter incompressibility coefficients in Table~\ref{tab1}.
Remarkably, the gaps between $L-\Delta{R}_{\mathrm{np}}^{48}$ and $L-\Delta{R}_{\mathrm{np}}^{208}$ increase with increasing incompressibility coefficients from $K=220$ MeV to 240 MeV.

\begin{table}[htbp!]
\centering
\caption{Slope parameters $L$ induced from the NSTs of $^{48}$Ca and $^{208}$Pb are shown by the classified isoscalar incompressibility coefficients. The systematic uncertainties are presented in the parenthesis.}\label{tab1}
\begin{tabular}{ccc}
\hline
\hline
~~~$K$~(MeV)& ~~~$L-\Delta{R}_{\mathrm{np}}^{48}$~(MeV)~~~   &  ~~$L-\Delta{R}_{\mathrm{np}}^{208}$~(MeV)~~\\
\hline
220 & $0\sim60.96$~(3.08)  &  $62.94\sim136.65$~(1.70)  \\
230 & $0\sim57.64$~(2.87)  &  $74.05\sim155.99$~(1.64)  \\
240 & $0\sim52.78$~(2.54)  &  $81.35\sim168.01$~(1.33)  \\
\hline\hline
\end{tabular}
\end{table}
Nuclear matter properties consisting of isovector and isoscalar components are correlated with each other.
Ref.~\cite{PhysRevC.73.044320} suggests that there is no clear correlation between the incompressibility $K$ and NSE, and between the slope of the NSE and incompressibility $K$.
The correlations between $K$ and the isovector parameters are generally weaker than those between the NST and NSE coefficients~\cite{PhysRevC.69.024318,PhysRevC.104.054324}.
As seen in Fig.~\ref{fig1}~(b), the increasing incompressibility coefficient $K$ influences the determination of the covered range of the slope parameter $L$.
Table~\ref{tab1} shows that the gap between $L-\Delta{R}_{\mathrm{np}}^{48}$ and $L-\Delta{R}_{\mathrm{np}}^{208}$ is smaller  than the theoretical uncertainty when the nuclear incompressibility is $K=220$ MeV.
This is instructive for calibrating new sets of Skyrme parameters for reproducing various nuclear matter properties as auxiliary conditions.

\begin{figure}[htbp]
  \centering
    \includegraphics[width=1.1\linewidth]{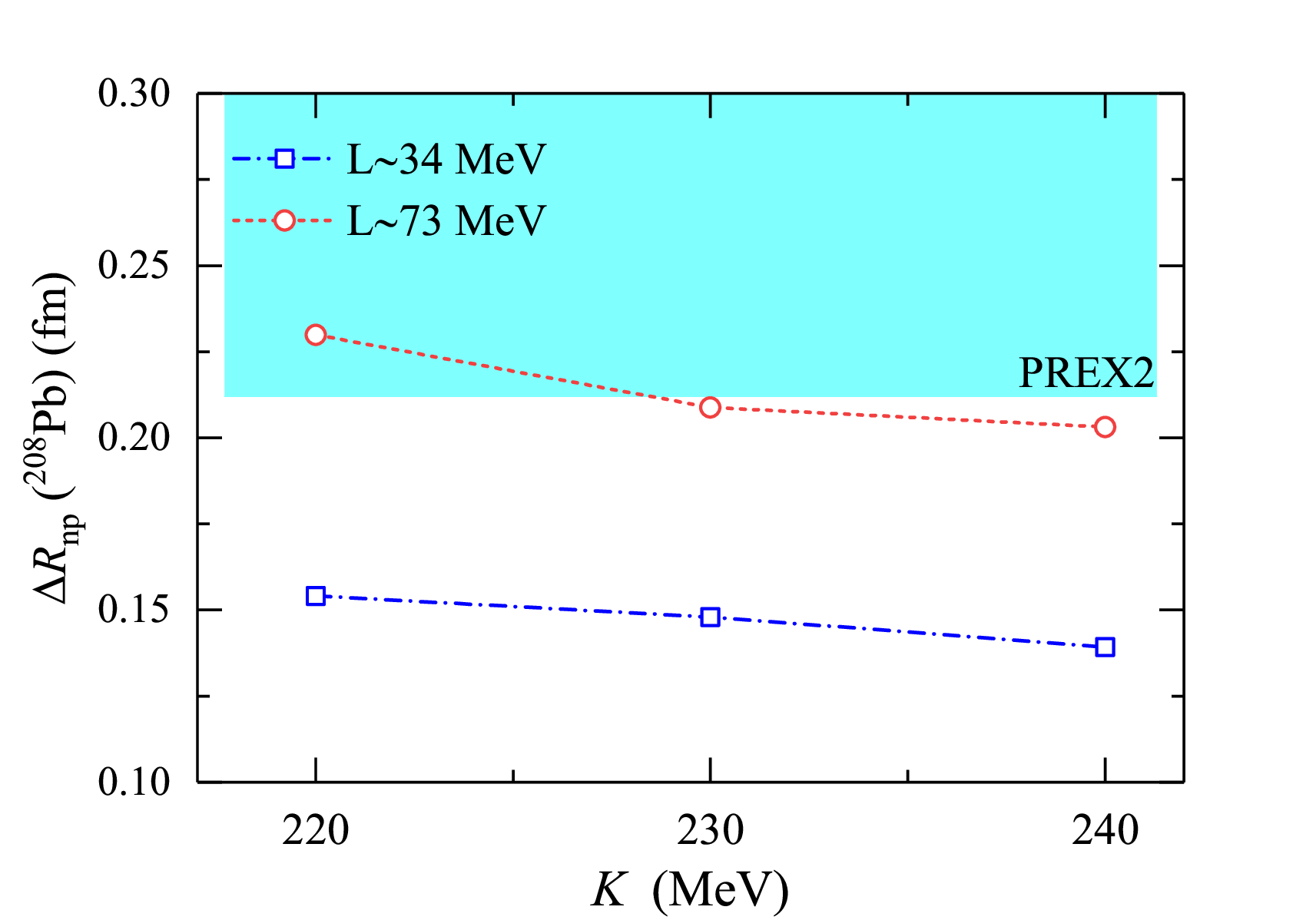}
   \caption{(Color online) Neutron skin thickness of $^{208}$Pb as a function of incompressibility coefficient $K$ at saturation density $\rho_{0}$.}\label{fig2}
\end{figure}
To facilitate the influence of incompressibility coefficient on determining the slope parameter $L$, the ``data-to-data" relations between the NST of $^{208}$Pb and the incompressibility coefficients $K$ are presented in Fig.~\ref{fig2}. Here, the slope parameters of the NSE were chosen to be approximately $L=34$ MeV and $L=73$ MeV. From this figure, in can be seen that the NST of $^{208}$Pb decreases with increasing incompressibility coefficient. This further demonstrates that the isoscalar compression modulus should be appropriately considered in the calibration protocol.

In our calculations, the upper limits of $L$ are gradually overestimated as the increasing incompressibility coefficients $K$ increased.
Combined with the latest PREX2 experiment, the result extracted from the relativistic EDFs leads to a covered range of $L = 106\pm37$ MeV~\cite{PhysRevLett.126.172503}.
The induced slope parameter $L$ is more consistent with that obtained when the incompressibility coefficient is $K=220$ MeV.

In Refs.~\cite{PhysRevC.88.011301,PhysRevLett.119.122502,PhysRevResearch.2.022035}, the highly linear correlation between the slope parameter $L$ and the differences of charge radii of mirror-partner nuclei $\Delta{R_{\mathrm{ch}}}$ was demonstrated.
The nuclear charge radius of $^{54}$Ni has been determined using the collinear laser spectroscopy~\cite{PhysRevLett.127.182503}. By combining the charge radii of the mirror-pair nuclei $^{54}$Fe, the deduced slope parameter covers the interval range $21\leq{L}\leq88$ MeV.
A recent study suggested that the upper or lower limits of $L$ may be constrained if precise data on the mirror charge radii of $^{44}$Cr-$^{44}$Ca and $^{46}$Fe-$^{46}$Ca are selected~\cite{PhysRevC.107.034319}.
In all of these studies, isoscalar nuclear matter properties were not considered.
In fact, the value deduced from the relativistic and non-relativistic Skyrme EDFs with identical incompressibility coefficients $K=230$ MeV gives a narrow range of $22.50\leq{L}\leq51.55$ MeV~\cite{An:2023ahu}.
This is in agreement with the results in Ref.~\cite{Konig:2023rwe} where a soft EoS is obtained, for example, $L\leq60$ MeV.

\begin{figure}[htbp]
  \centering
    \includegraphics[width=1.1\linewidth]{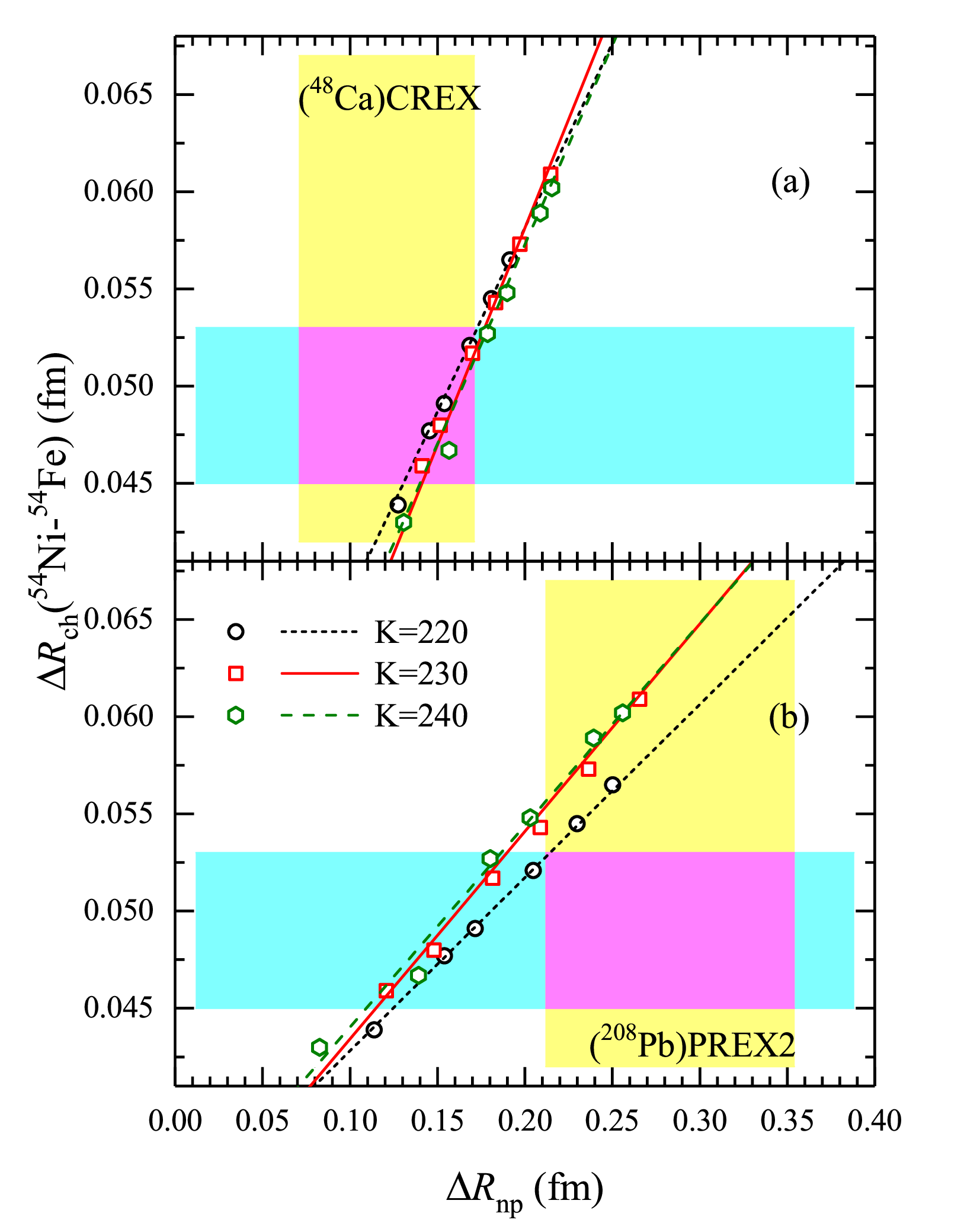}
   \caption{(Color online) $\Delta{R}_{\mathrm{ch}}$ of the mirror-pair nuclei $^{54}$Ni-$^{54}$Fe as a function of the neutron skin thickness of $^{48}$Ca (a) and $^{208}$Pb (b). The experimental constraints are shown as a horizontal light-blue band. The open markers are the results of Skyrme-EDF calculations. The corresponding lines indicate theoretical linear fits.}\label{fig3}
\end{figure}

In atomic nuclei, the NST is regarded as a perfect signal for describing the isovector property, and is highly correlated with the slope parameter of the NSE.
The difference in the charge radii of the mirror-pair nuclei and the slope of the NSE exhibits a highly linear relationship~\cite{PhysRevC.97.014314,PhysRevLett.130.032501,PhysRevC.108.015802}.
To facilitate the influence of the isoscalar properties on determining the EoS of nuclear matter, the data-to-data relations between the difference in charge radii $\Delta{R}_{\mathrm{ch}}$ of the mirror-pair nuclei $^{54}$Ni-$^{54}$Fe and the NSTs of $^{48}$Ca and $^{208}$Pb are shown in Fig.~\ref{fig3}.
Notably, highly linear correlations between $\Delta{R}_{\mathrm{ch}}$ and the NSTs of $^{48}$Ca and $^{208}$Pb are observed.

In Fig.~\ref{fig3}~(a), the linear functions fit the experimental data well across various incompressibility coefficients $K$, that is, the slope parameter can be constrained concurrently through the calculated NST of $^{48}$Ca and the $\Delta{R}_{\mathrm{ch}}$ of mirror-pair nuclei $^{54}$Ni-$^{54}$Fe. However, as shown in Fig.~\ref{fig3}~(b), the fitting lines deviate from the cross-over region between $\Delta{R}_{\mathrm{ch}}$ and the NST of $^{208}$Pb except for $K=220$ MeV. Although the linear function captures a relatively narrow region, this further demonstrates the need to extract valid information about the nuclear EoS by considering the isoscalar components in the calibration procedure.

The Coulomb term does not contribute to infinite nuclear matter calculations, in which the NSE plays an essential role in determining the evolution of isospin asymmetry components.
However, in atomic nuclei, the actual proton and neutron density distributions are mostly dominated by the degree of isospin asymmetry and Coulomb forces.
It is evident that the competition between the Coulomb interaction and the NSE is related to the stability of dripline nuclei against nucleon emission~\cite{RocaMaza:2018ujj,PhysRevC.82.027301}.
The NST is associated with the symmetry energy and significantly influenced by the NSE, which corresponds to the EoS of neutron-rich matter.
Meanwhile, a strong high liner correlation between the slope parameter $L$ and the difference in the charge radii of mirror-pair nuclei is evident~\cite{PhysRevC.88.011301,PhysRevLett.119.122502,PhysRevResearch.2.022035,PhysRevLett.127.182503,PhysRevC.107.034319,An:2023ahu}.
As shown in Fig.~\ref{fig3}, this highly linear correlation extends to the NST and the difference in charge radii of mirror-pair nuclei, owing to the isospin-symmetry breaking~\cite{PhysRevC.108.015802}.

\section{Summary and outlook}\label{sec3}
As is well known, the Skyrme parameters can be characterized analytically by the isoscalar and isovector nuclear matter properties of the Hamiltonian density.
More effective statistical methods have also been used to discuss the theoretical uncertainties~\cite{PhysRevC.93.051303,Balliet_2021,PhysRevC.105.L021301}.
In this study, we reviewed the influence of nuclear matter incompressibility on the determination of the slope parameter of symmetry energy $L$.
The NSTs of $^{48}$Ca and $^{208}$Pb were calculated using Skyrme EDFs.
The slope parameter $L$ deduced from $^{208}$Pb is sensitive to the incompressibility coefficients, whereas that for $^{48}$Ca is not.
A continuous range of $L$ can be obtained if the nuclear matter is incompressible at $K=220$ MeV.
This is in agreement with that in Ref.~\cite{PhysRevC.104.054324} where the nuclear matter incompressibility covers the interval range of $K=223^{+7}_{-8}$ MeV.
This implies that the isoscalar components should be considered when determining the slope parameter $L$.
In addition, it is desirable to review the influence of the incompressibility coefficient $K$ on the determination of the slope parameter $L$ within the framework of relativistic EDFs.

The nuclear symmetry energy can be obtained using different methods and models~\cite{Giuliani2023,FURNSTAHL200285,PhysRevC.93.064303,GAIDAROV2020122061,Cheng_2023,Caoantigdr,Caoantigdr1,Cao08,Cao041,LiuMin2011,
Fangdeqing,GAUTAM2024122832,fang2023,Gao2023,Xu_2021,He2023}.
The precise determination of the slope parameter $L$ is related to various quantities such as the charge changing cross section~\cite{XU2022137333,ZHAO2023138269}, sub-barrier fusion cross-section, and astrophysical $S$-factor in asymmetric nuclei~\cite{ghosh2023neutron}.
Generally, the proton and neutron density distributions are mutually determined by the isospin asymmetry and Coulomb force.
The isospin symmetry breaking effect influences the determination of the charge density distributions~\cite{DONG2019133,PhysRevC.105.L021304,PhysRevC.107.064302,SENG2023137654}.
Thus, more accurate descriptions of NST and charge radii are required.
In addition, the curvature of the symmetry energy $K_\mathrm{sym}$~\cite{PhysRevC.69.041301} and three-body interactions in the
 Skyrme forces~\cite{PhysRevC.94.064326} may also influence the determination of the neutron skin.

\section{Acknowledgements}
This work is supported partly by the National Key R$\&$D Program of China under Grant No. 2023YFA1606401 and the National Natural Science Foundation of China under Grants No. 12135004, No. 11635003, No. 11961141004 and No. 12047513. L.-G. C. is grateful for the support of the National Natural Science Foundation of China under Grants No. 12275025, No. 11975096 and the Fundamental Research Funds for the Central Universities (2020NTST06)¡£


\begin{thebibliography}{0}%
\makeatletter
\providecommand \@ifxundefined [1]{%
 \@ifx{#1\undefined}
}%
\providecommand \@ifnum [1]{%
 \ifnum #1\expandafter \@firstoftwo
 \else \expandafter \@secondoftwo
 \fi
}%
\providecommand \@ifx [1]{%
 \ifx #1\expandafter \@firstoftwo
 \else \expandafter \@secondoftwo
 \fi
}%
\providecommand \natexlab [1]{#1}%
\providecommand \enquote  [1]{``#1''}%
\providecommand \bibnamefont  [1]{#1}%
\providecommand \bibfnamefont [1]{#1}%
\providecommand \citenamefont [1]{#1}%
\providecommand \href@noop [0]{\@secondoftwo}%
\providecommand \href [0]{\begingroup \@sanitize@url \@href}%
\providecommand \@href[1]{\@@startlink{#1}\@@href}%
\providecommand \@@href[1]{\endgroup#1\@@endlink}%
\providecommand \@sanitize@url [0]{\catcode `\\12\catcode `\$12\catcode `\&12\catcode `\#12\catcode `\^12\catcode `\_12\catcode `\%12\relax}%
\providecommand \@@startlink[1]{}%
\providecommand \@@endlink[0]{}%
\providecommand \url  [0]{\begingroup\@sanitize@url \@url }%
\providecommand \@url [1]{\endgroup\@href {#1}{\urlprefix }}%
\providecommand \urlprefix  [0]{URL }%
\providecommand \Eprint [0]{\href }%
\providecommand \doibase [0]{http://dx.doi.org/}%
\providecommand \selectlanguage [0]{\@gobble}%
\providecommand \bibinfo  [0]{\@secondoftwo}%
\providecommand \bibfield  [0]{\@secondoftwo}%
\providecommand \translation [1]{[#1]}%
\providecommand \BibitemOpen [0]{}%
\providecommand \bibitemStop [0]{}%
\providecommand \bibitemNoStop [0]{.\EOS\space}%
\providecommand \EOS [0]{\spacefactor3000\relax}%
\providecommand \BibitemShut  [1]{\csname bibitem#1\endcsname}%
\let\auto@bib@innerbib\@empty
\end{thebibliography}%


\begin{thebibliography}{99}

\bibitem{PhysRevLett.86.5647} C. J. Horowitz and J. Piekarewicz, \textit{Neutron Star Structure and the Neutron Radius of $^{208}\mathrm{Pb}$}.
Phys. Rev. Lett. 86, 5647 (2001).
\href{https://link.aps.org/doi/10.1103/PhysRevLett.86.5647}{https://doi.org/10.1103/PhysRevLett.86.5647}

\bibitem{Li2019} B.-A. Li, P. G. Krastev, D.-H. Wen $et~al$., \textit{Towards understanding astrophysical effects of nuclear symmetry energy}. Eur. Phys. J. A 55, 117 (2019).
\href{https://doi.org/10.1140/epja/i2019-12780-8}{https://doi.org/10.1140/epja/i2019-12780-8}

\bibitem{Steiner:2004fi}  A. W. Steiner, M. Prakash, J. M. Lattimer $et~al$., \textit{Isospin asymmetry in nuclei and neutron stars}. Phys. Rept. 411, 325 (2005).
\href{https://www.sciencedirect.com/science/article/pii/S0370157305001043}{https://doi.org/10.1016/j.physrep.2005.02.004}

\bibitem{LATTIMER2007109} J. M. Lattimer and M. Prakash, \textit{Neutron star observations: Prognosis for equation of state constraints}. Phys. Rept. 442, 109 (2007), the Hans Bethe Centennial Volume 1906-2006.
\href{https://www.sciencedirect.com/science/article/pii/S0370157307000452}{https://doi.org/10.1016/j.physrep.2007.02.003}

\bibitem{Xiachengjun} J. F. Xu, C. J. Xia, Z. Y. Lu $et~al$., \textit{Symmetry energy of strange quark matter and tidal deformability of strange quark stars}. Nucl. Sci. Tech. 33, 143 (2022).
\href{https://doi.org/10.1007/s41365-022-01130-x}{https://doi.org/10.1007/s41365-022-01130-x}

\bibitem{Li:2008gp}  B.-A. Li, L.-W. Chen, and C. M. Ko, \textit{Recent Progress and New Challenges in Isospin Physics with Heavy-Ion Reactions}. Phys. Rept. 464, 113 (2008).
\href{https://www.sciencedirect.com/science/article/pii/S0370157308001269}{https://doi.org/10.1016/j.physrep.2008.04.005}

\bibitem{Li:2014oda} B.-A. Li, A. Ramos, G. Verde $et~al$., \textit{Topical issue on nuclear symmetry energy}. Eur. Phys. J. A 50, 9 (2014).
\href{https://doi.org/10.1140/epja/i2014-14009-x}{https://doi.org/10.1140/epja/i2014-14009-x}

\bibitem{Hu:2020ujf}  J.-N. Hu, S.-S. Bao, Y. Zhang $et~al$., \textit{Effects of symmetry energy on the radius and tidal deformability of neutron stars in the relativistic mean-field model}. Prog. Theor. Exp. Phys. 2020, 043D01 (2020).
\href{https://doi.org/10.1093/ptep/ptaa016}{https://doi.org/10.1093/ptep/ptaa016}

\bibitem{PhysRevC.90.064317} Z. Zhang and L.-W. Chen, \textit{Constraining the density slope of nuclear symmetry energy at subsaturation densities using electric dipole polarizability in $^{208}\mathrm{Pb}$}. Phys. Rev. C 90, 064317 (2014).
\href{https://link.aps.org/doi/10.1103/PhysRevC.90.064317}{https://doi.org/10.1103/PhysRevC.90.064317}

\bibitem{PhysRevC.101.034303} Y.-X. Zhang, M. Liu, and C.-J Xia $et~al$., \textit{Constraints on the symmetry energy and its associated parameters from nuclei to neutron stars}. Phys. Rev. C 101, 034303 (2020).
\href{https://link.aps.org/doi/10.1103/PhysRevC.101.034303}{https://doi.org/10.1103/PhysRevC.101.034303}

\bibitem{Liu_2018} J. Liu, Z.-Z. Ren and C. Xu, \textit{Combining the modified Skyrme-like model and the local density approximation to determine the symmetry energy of nuclear matter}. J. Phys. G 45, 075103 (2018).
\href{https://dx.doi.org/10.1088/1361-6471/aac78f}{https://doi.org/10.1088/1361-6471/aac78f}

\bibitem{PhysRevC.108.L021303} S. Yang, R.-J. Li and C. Xu, \textit{$\ensuremath{\alpha}$ clustering in nuclei and its impact on the nuclear symmetry energy}. Phys. Rev. C 108, L021303 (2023).
\href{https://link.aps.org/doi/10.1103/PhysRevC.108.L021303}{https://doi.org/10.1103/PhysRevC.108.L021303}

\bibitem{PhysRevLett.77.3963}  J. Meng and P. Ring, \textit{Relativistic Hartree-Bogoliubov Description of the Neutron Halo in $^{11}\mathrm{Li}$}. Phys. Rev. Lett. 77, 3963 (1996).
\href{https://link.aps.org/doi/10.1103/PhysRevLett.77.3963}{https://doi.org/10.1103/PhysRevLett.77.3963}

\bibitem{PhysRevC.82.011301} S.-G. Zhou, J. Meng, P. Ring $et~al$., \textit{Neutron halo in deformed nuclei}. Phys. Rev. C 82, 011301(R) (2010).
\href{https://link.aps.org/doi/10.1103/PhysRevC.82.011301}{https://doi.org/10.1103/PhysRevC.82.011301}

\bibitem{PhysRevC.102.044313}  X.-N. Cao, K.-M. Ding, M. Shi $et~al$., \textit{Exploration of the exotic structure in $\mathrm{Ce}$ isotopes by the relativistic point-coupling model combined with complex momentum representation}. Phys. Rev. C 102, 044313 (2020).
\href{https://link.aps.org/doi/10.1103/PhysRevC.102.044313}{https://doi.org/10.1103/PhysRevC.102.044313}

\bibitem{Brown20005296} B. A. Brown, \textit{Neutron radii in nuclei and the neutron equation of state}. Phys. Rev. Lett. 85, 5296 (2000).
\href{https://link.aps.org/doi/10.1103/PhysRevLett.85.5296}{https://doi.org/10.1103/PhysRevLett.85.5296}

\bibitem{PhysRevC.64.027302}  S. Typel and B. A. Brown, \textit{Neutron radii and the neutron equation of state in relativistic models}. Phys. Rev. C 64, 027302 (2001).
\href{https://link.aps.org/doi/10.1103/PhysRevC.64.027302}{https://doi.org/10.1103/PhysRevC.64.027302}

\bibitem{PhysRevC.69.024318} S. Yoshida and H. Sagawa, \textit{Neutron skin thickness and equation of state in asymmetric nuclear matter}. Phys. Rev. C 69, 024318 (2004).
\href{https://link.aps.org/doi/10.1103/PhysRevC.69.024318}{https://doi.org/10.1103/PhysRevC.69.024318}

\bibitem{PhysRevC.72.064309}  L.-W. Chen, C. M. Ko, and B.-A. Li, \textit{Nuclear matter symmetry energy and the neutron skin thickness of heavy nuclei}. Phys. Rev. C 72, 064309 (2005).
\href{https://link.aps.org/doi/10.1103/PhysRevC.72.064309}{https://doi.org/10.1103/PhysRevC.72.064309}

\bibitem{PhysRevLett.106.252501} X. Roca-Maza, M. Centelles, X. Vi\~{n}as $et~al$., \textit{Neutron Skin of $^{208}\mathrm{Pb}$, Nuclear Symmetry Energy, and the Parity Radius Experiment}. Phys. Rev. Lett. 106, 252501 (2011).
\href{https://link.aps.org/doi/10.1103/PhysRevLett.106.252501}{https://doi.org/10.1103/PhysRevLett.106.252501}

\bibitem{PhysRevC.80.024316}  M. Warda, X. Vi\~{n}as, X. Roca-Maza $et~al$., \textit{Neutron skin thickness in the droplet model with surface width dependence: Indications of softness of the nuclear symmetry energy}. Phys. Rev. C 80, 024316 (2009).
\href{https://link.aps.org/doi/10.1103/PhysRevC.80.024316}{https://doi.org/10.1103/PhysRevC.80.024316}

\bibitem{PhysRevC.84.034316}  M. K. Gaidarov, A. N. Antonov, P. Sarriguren $et~al$., \textit{Surface properties of neutron-rich exotic nuclei: A source for studying the nuclear symmetry energy}. Phys. Rev. C 84, 034316 (2011).
\href{https://link.aps.org/doi/10.1103/PhysRevC.84.034316}{https://doi.org/10.1103/PhysRevC.84.034316}

\bibitem{PhysRevLett.102.122502} M. Centelles, X. Roca-Maza, X. Vi\~{n}as $et~al$., \textit{Nuclear Symmetry Energy Probed by Neutron Skin Thickness of Nuclei}. Phys. Rev. Lett. 102, 122502 (2009).
\href{https://link.aps.org/doi/10.1103/PhysRevLett.102.122502}{https://doi.org/10.1103/PhysRevLett.102.122502}

\bibitem{PhysRevC.81.051303}  P.-G. Reinhard and W. Nazarewicz, \textit{Information content of a new observable: The case of the nuclear neutron skin}. Phys. Rev. C 81, 051303(R) (2010).
\href{https://link.aps.org/doi/10.1103/PhysRevC.81.051303}{https://doi.org/10.1103/PhysRevC.81.051303}

\bibitem{PhysRevLett.109.262501} B. K. Agrawal, J. N. De, and S. K. Samaddar, \textit{Determining the Density Content of Symmetry Energy and Neutron Skin: An Empirical Approach}. Phys. Rev. Lett. 109, 262501 (2012).
\href{https://link.aps.org/doi/10.1103/PhysRevLett.109.262501}{https://doi.org/10.1103/PhysRevLett.109.262501}

\bibitem{PhysRevC.87.051306} B. K. Agrawal, J. N. De, S. K. Samaddar $et~al$., \textit{Constraining the density dependence of the symmetry energy from nuclear masses}. Phys. Rev. C 87, 051306(R) (2013).
\href{https://link.aps.org/doi/10.1103/PhysRevC.87.051306}{https://doi.org/10.1103/PhysRevC.87.051306}

\bibitem{PhysRevC.87.034327} N. Wang, L. Ou, and M. Liu, \textit{Nuclear symmetry energy from the Fermi-energy difference in nuclei}. Phys. Rev. C 87, 034327 (2013).
\href{https://link.aps.org/doi/10.1103/PhysRevC.87.034327}{https://doi.org/10.1103/PhysRevC.87.034327}

\bibitem{PhysRevC.93.051303} P.-G. Reinhard and W. Nazarewicz, \textit{Nuclear charge and neutron radii and nuclear matter: Trend analysis in Skyrme density-functional-theory approach}. Phys. Rev. C 93, 051303(R) (2016).
\href{https://link.aps.org/doi/10.1103/PhysRevC.93.051303}{https://doi.org/10.1103/PhysRevC.93.051303}

\bibitem{ZHANG2013234} Z. Zhang and L.-W. Chen, \textit{Constraining the symmetry energy at subsaturation densities using isotope binding energy difference and neutron skin thickness}. Phys. Lett. B 726, 234 (2013).
\href{https://www.sciencedirect.com/science/article/pii/S037026931300628X}{https://doi.org/10.1016/j.physletb.2013.08.002}

\bibitem{PhysRevC.102.044316}  J. Xu, W.-J. Xie, and B.-A. Li, \textit{Bayesian inference of nuclear symmetry energy from measured and imagined neutron skin thickness in $^{116,118,120,122,124,130,132}\mathrm{Sn}, ^{208}\mathrm{Pb}$, and $^{48}\mathrm{Ca}$}. Phys. Rev. C 102, 044316 (2020).
\href{https://link.aps.org/doi/10.1103/PhysRevC.102.044316}{https://doi.org/10.1103/PhysRevC.102.044316}

\bibitem{particles6010003} J. M. Lattimer, \textit{Constraints on Nuclear Symmetry Energy Parameters}. Particles 6, 30 (2023).
\href{https://www.mdpi.com/2571-712X/6/1/3}{https://doi.org/10.3390/particles6010003}

\bibitem{PhysRevC.91.034315} J.-M. Dong, W. Zuo and J.-Z. Gu, \textit{Constraints on neutron skin thickness in $^{208}\mathrm{Pb}$ and density-dependent symmetry energy}. Phys. Rev. C 91, 034315 (2015).
\href{https://link.aps.org/doi/10.1103/PhysRevC.91.034315}{https://doi.org/10.1103/PhysRevC.91.034315}

\bibitem{PhysRevC.90.064310} C. Xu, Z.-Z. Ren and J. Liu, \textit{Attempt to link the neutron skin thickness of $^{208}\mathrm{Pb}$ with the symmetry energy through cluster radioactivity}. Phys. Rev. C 90, 064310 (2014).
\href{https://link.aps.org/doi/10.1103/PhysRevC.90.064310}{https://doi.org/10.1103/PhysRevC.90.064310}

\bibitem{chen2006} L.-W. Chen, C. M. Ko and B.-A. Li, \textit{Constraining the Skyrme effective interactions and the neutron skin thickness of nuclei using isospin diffusion data from heavy ion collisions}. Int. J. Mod. Phys. E 15, 1385-1395 (2006).
\href{https://doi.org/10.1142/S0218301306004946}{https://doi.org/10.1142/S0218301306004946}

\bibitem{PhysRevC.97.034318} J.-M. Dong, L.-J. Wang, W. Zuo $et~al$., \textit{Constraints on Coulomb energy, neutron skin thickness in $^{208}\mathrm{Pb}$, and symmetry energy}. Phys. Rev. C 97, 034318 (2018).
\href{https://link.aps.org/doi/10.1103/PhysRevC.97.034318}{https://doi.org/10.1103/PhysRevC.97.034318}

\bibitem{PhysRevLett.126.172502} D. Adhikari, H. Albataineh, D. Androic, $et~al$. (PREX
Collaboration), \textit{Accurate Determination of the Neutron Skin Thickness of $^{208}\mathrm{Pb}$ through Parity-Violation in Electron Scattering}. Phys. Rev. Lett. 126, 172502 (2021).
\href{https://link.aps.org/doi/10.1103/PhysRevLett.126.172502}{https://doi.org/10.1103/PhysRevLett.126.172502}


\bibitem{PhysRevLett.129.042501} D. Adhikari, H. Albataineh, D. Androic, $et~al$. (CREX Collaboration), \textit{Precision Determination of the Neutral Weak Form Factor of $^{48}\mathrm{Ca}$}. Phys. Rev. Lett. 129, 042501 (2022).
\href{https://link.aps.org/doi/10.1103/PhysRevLett.129.042501}{https://doi.org/10.1103/PhysRevLett.129.042501}


\bibitem{PhysRevLett.118.252501} J. Birkhan, M. Miorelli, S. Bacca, $et~al$., \textit{Electric Dipole Polarizability of $^{48}\mathrm{Ca}$ and Implications for the Neutron Skin}. Phys. Rev. Lett. 118, 252501 (2017).
\href{https://link.aps.org/doi/10.1103/PhysRevLett.118.252501}{https://doi.org/10.1103/PhysRevLett.118.252501}


\bibitem{PhysRevLett.107.062502}  A. Tamii, I. Poltoratska, P. von Neumann-Cosel $et~al$., \textit{Complete Electric Dipole Response and the Neutron Skin in $^{208}\mathrm{Pb}$}. Phys. Rev. Lett. 107, 062502 (2011).
\href{https://link.aps.org/doi/10.1103/PhysRevLett.107.062502}{https://doi.org/10.1103/PhysRevLett.107.062502}


\bibitem{PhysRevLett.127.192701}  R. Essick, I. Tews, P. Landry $et~al$., \textit{Astrophysical Constraints on the Symmetry Energy and the Neutron Skin of $^{208}\mathrm{Pb}$ with Minimal Modeling Assumptions}. Phys. Rev. Lett. 127, 192701 (2021).
\href{https://link.aps.org/doi/10.1103/PhysRevLett.127.192701}{https://doi.org/10.1103/PhysRevLett.127.192701}

\bibitem{PhysRevLett.127.232501} P.-G. Reinhard, X. Roca-Maza, and W. Nazarewicz, \textit{Information Content of the Parity-Violating Asymmetry in $^{208}\mathrm{Pb}$}. Phys. Rev. Lett. 127, 232501 (2021).
\href{https://link.aps.org/doi/10.1103/PhysRevLett.127.232501}{https://doi.org/10.1103/PhysRevLett.127.232501}


\bibitem{PhysRevC.107.035802} H. Sotani and T. Naito, \textit{Empirical neutron star mass formula based on experimental observables}. Phys. Rev. C 107, 035802 (2023).
\href{https://link.aps.org/doi/10.1103/PhysRevC.107.035802}{https://doi.org/10.1103/PhysRevC.107.035802}


\bibitem{TAGAMI2022106037} S. Tagami, T. Wakasa, and M. Yahiro, \textit{Slope parameters determined from CREX and PREX2}. Res. Phys. 43, 106037 (2022).
\href{https://www.sciencedirect.com/science/article/pii/S2211379722006519}{https://doi.org/10.1016/j.rinp.2022.106037}

\bibitem{PhysRevLett.129.232501}  P.-G. Reinhard, X. Roca-Maza, and W. Nazarewicz, \textit{Combined Theoretical Analysis of the Parity-Violating Asymmetry for $^{48}\mathrm{Ca}$ and $^{208}\mathrm{Pb}$}. Phys. Rev. Lett. 129, 232501 (2022).
\href{https://link.aps.org/doi/10.1103/PhysRevLett.129.232501}{https://doi.org/10.1103/PhysRevLett.129.232501}

\bibitem{Zhang:2022bni} Z. Zhang and L.-W. Chen, \textit{Bayesian inference of the symmetry energy and the neutron skin in $^{48}\mathrm{Ca}$ and $^{208}\mathrm{Pb}$ from CREX and PREX-2}. Phys. Rev. C 108, 024317 (2023).
\href{https://link.aps.org/doi/10.1103/PhysRevC.108.024317}{https://doi.org/10.1103/PhysRevC.108.024317}

\bibitem{Miyatsu:2023lki} T. Miyatsu, M.-K. Cheoun, K. Kim $et~al$., \textit{Can the PREX-2 and CREX results be understood by relativistic mean-field models with the astrophysical constraints?}. Phys. Lett. B 843, 138013 (2023).
\href{https://www.sciencedirect.com/science/article/pii/S0370269323003477}{https://doi.org/10.1016/j.physletb.2023.138013}

\bibitem{YUKSEL2023137622} E. Y\"{u}ksel and N. Paar, \textit{Implications of parity-violating electron scattering experiments on 48Ca (CREX) and 208Pb (PREX-II) for nuclear energy density functionals}. Phys. Lett. B 836, 137622 (2023).
\href{https://www.sciencedirect.com/science/article/pii/S0370269322007560}{https://doi.org/10.1016/j.physletb.2022.137622}


\bibitem{PhysRevC.107.015801}  C. Mondal and F. Gulminelli, \textit{Nucleonic metamodeling in light of multimessenger, PREX-II, and CREX data}. Phys. Rev. C 107, 015801 (2023).
\href{https://link.aps.org/doi/10.1103/PhysRevC.107.015801}{https://doi.org/10.1103/PhysRevC.107.015801}

\bibitem{PhysRevC.96.065805} D. Chatterjee, F. Gulminelli, A. R. Raduta $et~al$., \textit{Constraints on the nuclear equation of state from nuclear masses and radii in a Thomas-Fermi meta-modeling approach}. Phys. Rev. C 96, 065805 (2017).
\href{https://link.aps.org/doi/10.1103/PhysRevC.96.065805}{https://doi.org/10.1103/PhysRevC.96.065805}

\bibitem{PhysRevC.73.044320} S. Yoshida and H. Sagawa, \textit{Isovector nuclear matter properties and neutron skin thickness}. Phys. Rev. C 73, 044320 (2006).
\href{https://link.aps.org/doi/10.1103/PhysRevC.73.044320}{https://doi.org/10.1103/PhysRevC.73.044320}

\bibitem{PhysRevC.70.024307} G. Col\`{o}, N. V. Giai, J. Meyer $et~al$., \textit{Microscopic determination of the nuclear incompressibility within the nonrelativistic framework}. Phys. Rev. C 70, 024307 (2004).
\href{https://link.aps.org/doi/10.1103/PhysRevC.70.024307}{https://doi.org/10.1103/PhysRevC.70.024307}

\bibitem{PhysRevC.86.054313} L.-G. Cao, H. Sagawa, and G. Col\`{o}, \textit{Microscopic study of the isoscalar giant monopole resonance in Cd, Sn, and Pb isotopes}. Phys. Rev. C 86, 054313 (2012).
\href{https://link.aps.org/doi/10.1103/PhysRevC.86.054313}{https://doi.org/10.1103/PhysRevC.86.054313}

\bibitem{PhysRevC.70.041604}  A. Ono, P. Danielewicz, W. A. Friedman $et~al$., \textit{Symmetry energy for fragmentation in dynamical nuclear collisions}. Phys. Rev. C 70, 041604(R) (2004).
\href{https://link.aps.org/doi/10.1103/PhysRevC.70.041604}{https://doi.org/10.1103/PhysRevC.70.041604}

\bibitem{PhysRevC.73.014313}  L.-G. Cao, U. Lombardo, C. W. Shen $et~al$., \textit{From Brueckner approach to Skyrme-type energy density functional}. Phys. Rev. C 73, 014313 (2006).
\href{https://link.aps.org/doi/10.1103/PhysRevC.73.014313}{https://doi.org/10.1103/PhysRevC.73.014313}

\bibitem{PhysRevC.104.054324} J. Xu, Z. Zhang, and B.-A. Li, \textit{Bayesian uncertainty quantification for nuclear matter incompressibility}. Phys. Rev. C 104, 054324 (2021).
\href{https://link.aps.org/doi/10.1103/PhysRevC.104.054324}{https://doi.org/10.1103/PhysRevC.104.054324}


\bibitem{RevModPhys.75.121} M. Bender, P.-H. Heenen, and P.-G. Reinhard, \textit{Self-consistent mean-field models for nuclear structure}. Rev. Mod. Phys. 75, 121 (2003).
\href{https://link.aps.org/doi/10.1103/RevModPhys.75.121}{https://doi.org/10.1103/RevModPhys.75.121}

\bibitem{PhysRevC.95.014316} Y. Zhang, Y. Chen, J. Meng $et~al$., \textit{Influence of pairing correlations on the radius of neutron-rich nuclei}. Phys. Rev. C 95, 014316 (2017).
\href{https://link.aps.org/doi/10.1103/PhysRevC.95.014316}{https://doi.org/10.1103/PhysRevC.95.014316}

\bibitem{PhysRevC.102.054312} Y. Zhang and X. Y. Qu, \textit{Effects of pairing correlation on the quasiparticle resonance in neutron-rich $\mathrm{Ca}$ isotopes}. Phys. Rev. C 102, 054312 (2020).
\href{https://link.aps.org/doi/10.1103/PhysRevC.102.054312}{https://doi.org/10.1103/PhysRevC.102.054312}

\bibitem{PhysRevC.90.024317} J. C. Pei, G. I. Fann, R. J. Harrison $et~al$., \textit{Adaptive multi-resolution 3D Hartree-Fock-Bogoliubov solver for nuclear structure}. Phys. Rev. C 90, 024317 (2014).
\href{https://link.aps.org/doi/10.1103/PhysRevC.90.024317}{https://doi.org/10.1103/PhysRevC.90.024317}

\bibitem{PhysRevC.102.014312} Q. Z. Chai, J. C. Pei, N. Fei $et~al$., \textit{Constraints on the neutron drip line with the newly observed $^{39}\mathrm{Na}$}. Phys. Rev. C 102, 014312 (2020).
\href{https://link.aps.org/doi/10.1103/PhysRevC.102.014312}{https://doi.org/10.1103/PhysRevC.102.014312}

\bibitem{WU2022136886} Z.-J. Wu, L. Guo, Z. Liu $et~al$., \textit{Production of proton-rich nuclei in the vicinity of 100Sn via multinucleon transfer reactions}. Phys. Lett. B 825, 136886 (2022).
\href{https://www.sciencedirect.com/science/article/pii/S037026932200020X}{https://doi.org/10.1016/j.physletb.2022.136886}

\bibitem{Caotens11} L.-G. Cao, H. Sagawa, and G. Col\`{o}, \textit{Effects of tensor correlations on low-lying collective states in finite nuclei}. Phys. Rev. C 83, 034324 (2011).
\href{https://link.aps.org/doi/10.1103/PhysRevC.83.034324}{https://doi.org/10.1103/PhysRevC.83.034324}

\bibitem{Wenpw14} P.-W. Wen, L.-G. Cao, J. Margueron $et~al$., \textit{Spin-isospin response in finite nuclei from an extended Skyrme interaction}. Phys. Rev. C 89, 044311 (2014).
\href{https://link.aps.org/doi/10.1103/PhysRevC.89.044311}{https://doi.org/10.1103/PhysRevC.89.044311}

\bibitem{PhysRevC.87.064311} E. Khan, N. Paar, D. Vretenar $et~al$., \textit{Incompressibility of finite fermionic systems: Stable and exotic atomic nuclei}. Phys. Rev. C 87, 064311 (2013).
\href{https://link.aps.org/doi/10.1103/PhysRevC.87.064311}{https://doi.org/10.1103/PhysRevC.87.064311}

\bibitem{Caoquench} L.-G. Cao, S.-S. Zhang, and H. Sagawa, \textit{Quenching factor of Gamow-Teller and spin dipole giant resonances}. Phys. Rev. C 100, 054324 (2019).
\href{https://link.aps.org/doi/10.1103/PhysRevC.100.054324}{https://doi.org/10.1103/PhysRevC.100.054324}

\bibitem{CHABANAT1997710} E. Chabanat, P. Bonche, P. Haensel $et~al$., \textit{A Skyrme parametrization from subnuclear to neutron star densities}. Nucl. Phys. A 627, 710 (1997).
\href{https://www.sciencedirect.com/science/article/pii/S0375947497005964}{https://doi.org/10.1016/S0375-9474(97)00596-4}

\bibitem{CHABANAT1998231} E. Chabanat, P. Bonche, P. Haensel $et~al$., \textit{A Skyrme parametrization from subnuclear to neutron star densities Part II. Nuclei far from stabilities}. Nucl. Phys. A 635, 231 (1998).
\href{https://www.sciencedirect.com/science/article/pii/S0375947498001808}{https://doi.org/10.1016/S0375-9474(98)00180-8}

\bibitem{PhysRevC.85.035201} M. Dutra, O. Louren\c{c}o, J. S. S\'{a} Martins $et~al$., \textit{Skyrme interaction and nuclear matter constraints}. Phys. Rev. C 85, 035201 (2012).
\href{https://link.aps.org/doi/10.1103/PhysRevC.85.035201}{https://doi.org/10.1103/PhysRevC.85.035201}

\bibitem{PhysRevC.82.024321} L.-W. Chen, C. M. Ko, B.-A. Li $et~al$., \textit{Density slope of the nuclear symmetry energy from the neutron skin thickness of heavy nuclei}. Phys. Rev. C 82, 024321 (2010).
\href{https://link.aps.org/doi/10.1103/PhysRevC.82.024321}{https://doi.org/10.1103/PhysRevC.82.024321}

\bibitem{PhysRevLett.82.691} D. H. Youngblood, H. L. Clark, and Y.-W. Lui, \textit{Incompressibility of Nuclear Matter from the Giant Monopole Resonance}. Phys. Rev. Lett. 82, 691 (1999).
\href{https://link.aps.org/doi/10.1103/PhysRevLett.82.691}{https://doi.org/10.1103/PhysRevLett.82.691}

\bibitem{PhysRevC.69.051301}  M. Uchida, H. Sakaguchi, M. Itoh $et~al$., \textit{Systematics of the bimodal isoscalar giant dipole resonance}. Phys. Rev. C 69, 051301(R) (2004).
\href{https://link.aps.org/doi/10.1103/PhysRevC.69.051301}{https://doi.org/10.1103/PhysRevC.69.051301}

\bibitem{PhysRevC.74.034302} W. M. Seif, \textit{$\ensuremath{\alpha}$ decay as a probe of nuclear incompressibility}. Phys. Rev. C 74, 034302 (2006).
\href{https://link.aps.org/doi/10.1103/PhysRevC.74.034302}{https://doi.org/10.1103/PhysRevC.74.034302}


\bibitem{Chen_2012} L.-W. Chen and J.-Z. Gu, \textit{Correlations between the nuclear breathing mode energy and properties of asymmetric nuclear matter}. J. Phys. G 39, 035104 (2012).
\href{https://dx.doi.org/10.1088/0954-3899/39/3/035104}{https://doi.org/10.1088/0954-3899/39/3/035104}

\bibitem{PhysRevC.94.052801} N. Alam, B. K. Agrawal, M. Fortin $et~al$., \textit{Strong correlations of neutron star radii with the slopes of nuclear matter incompressibility and symmetry energy at saturation}. Phys. Rev. C 94, 052801(R) (2016).
\href{https://link.aps.org/doi/10.1103/PhysRevC.94.052801}{https://doi.org/10.1103/PhysRevC.94.052801}

\bibitem{PhysRevC.104.055804} A. Kumar, H. C. Das, and S. K. Patra, \textit{Incompressibility and symmetry energy of a neutron star}. Phys. Rev. C 104, 055804 (2021).
\href{https://link.aps.org/doi/10.1103/PhysRevC.104.055804}{https://doi.org/10.1103/PhysRevC.104.055804}

\bibitem{PhysRevC.69.041301} J. Piekarewicz, \textit{Unmasking the nuclear matter equation of state}. Phys. Rev. C 69, 041301(R) (2004).
\href{https://link.aps.org/doi/10.1103/PhysRevC.69.041301}{https://doi.org/10.1103/PhysRevC.69.041301}

\bibitem{PhysRevLett.126.172503} B. T. Reed, F. J. Fattoyev, C. J. Horowitz $et~al$., \textit{Implications of PREX-2 on the Equation of State of Neutron-Rich Matter}. Phys. Rev. Lett. 126, 172503 (2021).
\href{https://link.aps.org/doi/10.1103/PhysRevLett.126.172503}{https://doi.org/10.1103/PhysRevLett.126.172503}

\bibitem{PhysRevC.88.011301} N. Wang and T. Li, \textit{Shell and isospin effects in nuclear charge radii}. Phys. Rev. C 88, 011301(R) (2013).
\href{https://link.aps.org/doi/10.1103/PhysRevC.88.011301}{https://doi.org/10.1103/PhysRevC.88.011301}

\bibitem{PhysRevLett.119.122502} B. A. Brown, \textit{Mirror Charge Radii and the Neutron Equation of State}. Phys. Rev. Lett. 119, 122502 (2017).
\href{https://link.aps.org/doi/10.1103/PhysRevLett.119.122502}{https://doi.org/10.1103/PhysRevLett.119.122502}

\bibitem{PhysRevResearch.2.022035}  B. A. Brown, K. Minamisono, J. Piekarewicz $et~al$., \textit{Implications of the $^{36}\mathrm{Ca}\ensuremath{-}^{36}\mathrm{S}$ and $^{38}\mathrm{Ca}\ensuremath{-}^{38}\mathrm{Ar}$ difference in mirror charge radii on the neutron matter equation of state}. Phys. Rev. Research 2, 022035(R) (2020).
\href{https://link.aps.org/doi/10.1103/PhysRevResearch.2.022035}{https://doi.org/10.1103/PhysRevResearch.2.022035}

\bibitem{PhysRevLett.127.182503}  S. V. Pineda, K. K\"{o}nig, D. M. Rossi $et~al$., \textit{Charge Radius of Neutron-Deficient $^{54}\mathrm{Ni}$ and Symmetry Energy Constraints Using the Difference in Mirror Pair Charge Radii}. Phys. Rev. Lett. 127, 182503 (2021).
\href{https://link.aps.org/doi/10.1103/PhysRevLett.127.182503}{https://doi.org/10.1103/PhysRevLett.127.182503}

\bibitem{PhysRevC.107.034319} Y. N. Huang, Z. Z. Li, and Y. F. Niu, \textit{Correlation between the difference of charge radii in mirror nuclei and the slope parameter of the symmetry energy}. Phys. Rev. C 107, 034319 (2023).
\href{https://link.aps.org/doi/10.1103/PhysRevC.107.034319}{https://doi.org/10.1103/PhysRevC.107.034319}

\bibitem{An:2023ahu} R. An, S. Sun, L.-G. Cao $et~al$., \textit{Constraining nuclear symmetry energy with the charge radii of mirror-pair nuclei}. Nucl. Sci. Tech. 34, 119 (2023).
\href{https://doi.org/10.1007/s41365-023-01269-1}{https://doi.org/10.1007/s41365-023-01269-1}

\bibitem{Konig:2023rwe} K. K\"{o}nig, J. C. Berengut, A. Borschevsky $et~al$., \textit{Nuclear charge radii of silicon isotopes}. arXiv:2309.02037 [nucl-ex], (2023).
\href{https://arxiv.org/abs/2309.02037}{https://doi.org/10.48550/arXiv.2309.02037}

\bibitem{PhysRevC.97.014314} J. Yang and J. Piekarewicz, \textit{Difference in proton radii of mirror nuclei as a possible surrogate for the neutron skin}. Phys. Rev. C 97, 014314 (2018).
\href{https://link.aps.org/doi/10.1103/PhysRevC.97.014314}{https://doi.org/10.1103/PhysRevC.97.014314}

\bibitem{PhysRevLett.130.032501} S. J. Novario, D. Lonardoni, S. Gandolfi $et~al$., \textit{Trends of Neutron Skins and Radii of Mirror Nuclei from First Principles}. Phys. Rev. Lett. 130, 032501 (2023).
\href{https://link.aps.org/doi/10.1103/PhysRevLett.130.032501}{https://doi.org/10.1103/PhysRevLett.130.032501}

\bibitem{PhysRevC.108.015802} P. Bano, S. P. Pattnaik, M. Centelles, $et~al$., \textit{Correlations between charge radii differences of mirror nuclei and stellar observables}. Phys. Rev. C 108, 015802 (2023).
\href{https://link.aps.org/doi/10.1103/PhysRevC.108.015802}{https://doi.org/10.1103/PhysRevC.108.015802}

\bibitem{RocaMaza:2018ujj} X. Roca-Maza and N. Paar, \textit{Nuclear equation of state from ground and collective excited state properties of nuclei}. Prog. Part. Nucl. Phys. 101, 96 (2018).
\href{https://www.sciencedirect.com/science/article/pii/S0146641018300334}{https://doi.org/10.1016/j.ppnp.2018.04.001}

\bibitem{PhysRevC.82.027301} K. Oyamatsu, K. Iida, and H. Koura, \textit{Neutron drip line and the equation of state of nuclear matter}. Phys. Rev. C 82, 027301 (2010).
\href{https://link.aps.org/doi/10.1103/PhysRevC.82.027301}{https://doi.org/10.1103/PhysRevC.82.027301}

\bibitem{Balliet_2021} L. E. Balliet, W. G. Newton, S. Cantu $et~al$., \textit{Prior Probability Distributions of Neutron Star Crust Models}. Astrophys. J. 918, 79 (2021).
\href{https://dx.doi.org/10.3847/1538-4357/ac06a4}{https://doi.org/10.3847/1538-4357/ac06a4}

\bibitem{PhysRevC.105.L021301} P.-G. Reinhard and W. Nazarewicz, \textit{Information content of the differences in the charge radii of mirror nuclei}. Phys. Rev. C 105, L021301 (2022).
\href{https://link.aps.org/doi/10.1103/PhysRevC.105.L021301}{https://doi.org/10.1103/PhysRevC.105.L021301}

\bibitem{Giuliani2023} P. Giuliani, K. Godbey, E. Bonilla $et~al$., \textit{Bayes goes fast: Uncertainty quantification for a covariant energy density functional emulated by the reduced basis method}. Front. Phys. 10 (2023).
\href{https://www.frontiersin.org/articles/10.3389/fphy.2022.1054524}{https://doi.org/10.3389/fphy.2022.1054524}

\bibitem{FURNSTAHL200285} R. Furnstahl, \textit{Neutron radii in mean-field models}.  Nucl. Phys. A 706, 85 (2002).
\href{https://www.sciencedirect.com/science/article/pii/S0375947402008679}{https://doi.org/10.1016/S0375-9474(02)00867-9}

\bibitem{PhysRevC.93.064303}  C. Mondal, B. K. Agrawal, M. Centelles $et~al$., \textit{Model dependence of the neutron-skin thickness on the symmetry energy}. Phys. Rev. C 93, 064303 (2016).
\href{https://link.aps.org/doi/10.1103/PhysRevC.93.064303}{https://doi.org/10.1103/PhysRevC.93.064303}

\bibitem{GAIDAROV2020122061}  M. Gaidarov, I. Moumene, A. Antonov $et~al$., \textit{Proton and neutron skins and symmetry energy of mirror nuclei}. Nucl. Phys. A 1004, 122061 (2020).
\href{https://www.sciencedirect.com/science/article/pii/S0375947420303717}{https://doi.org/10.1016/j.nuclphysa.2020.122061}

\bibitem{Cheng_2023} S.-H. Cheng, J. Wen, L.-G. Cao $et~al$., \textit{Neutron skin thickness of 90Zr and symmetry energy constrained by charge exchange spin-dipole excitations}. Chin. Phys. C 47, 024102 (2023).
\href{https://dx.doi.org/10.1088/1674-1137/aca38e}{https://doi.org/10.1088/1674-1137/aca38e}

\bibitem{Caoantigdr}  L.-G. Cao, X. Roca-Maza, G. Col\`{o} $et~al$., \textit{Constraints on the neutron skin and symmetry energy from the anti-analog giant dipole resonance in $^{208}\mathrm{Pb}$}. Phys. Rev. C 92, 034308 (2015).
\href{https://link.aps.org/doi/10.1103/PhysRevC.92.034308}{https://doi.org/10.1103/PhysRevC.92.034308}

\bibitem{Caoantigdr1}  X. Roca-Maza, L.-G. Cao, G. Col\`{o} $et~al$., \textit{Fully self-consistent study of charge-exchange resonances and the impact on the symmetry energy parameters}. Phys. Rev. C 94, 044313 (2016).
\href{https://link.aps.org/doi/10.1103/PhysRevC.94.044313}{https://doi.org/10.1103/PhysRevC.94.044313}

\bibitem{Cao08} L.-G. Cao and Z.-Y. Ma, \textit{Symmetry energy and isovector giant dipole resonance in finite nuclei}. Chin. Phys. Lett 25, 1625 (2008).
\href{https://doi.org/10.1088/0256-307x/25/5/028}{https://doi.org/10.1088/0256-307x/25/5/028}

\bibitem{Cao041} L.-G. Cao and Z.-Y. Ma, \textit{Soft dipole modes in neutron-rich Ni-isotopes in QRRPA}. Mod. Phys. Lett. A 19, 2845 (2004).
\href{https://doi.org/10.1142/S0217732304015233}{https://doi.org/10.1142/S0217732304015233}

\bibitem{LiuMin2011} M. Liu, Z.-X. Li, N. Wang $et~al$., \textit{Exploring nuclear symmetry energy with isospin dependence on neutron skin thickness of nuclei}. Chin. Phys. C 35, 629 (2011).
\href{https://dx.doi.org/10.1088/1674-1137/35/7/006}{https://doi.org/10.1088/1674-1137/35/7/006}

\bibitem{Fangdeqing} H. Yu, D.-Q. Fang, Y.-G. Ma, \textit{Investigation of the symmetry energy of nuclear matter using isospin-dependent quantum molecular dynamics}. Nucl. Sci. Tech. 31, 61 (2020).
\href{https://doi.org/10.1007/s41365-020-00766-x}{https://doi.org/10.1007/s41365-020-00766-x}

\bibitem{GAUTAM2024122832} S. Gautam, A. Venneti, S. Banik $et~al$., \textit{Estimation of the slope of nuclear symmetry energy via charge radii of mirror nuclei}. Nucl. Phys. A 1043, 122832 (2024).
\href{https://www.sciencedirect.com/science/article/pii/S0375947424000149}{https://doi.org/10.1016/j.nuclphysa.2024.122832}


\bibitem{fang2023} D.-Q. Fang, \textit{Neutron skin thickness and its effects in nuclear reactions}. Nucl. Tech. 46, 155 (2023).
\href{https://doi.org/10.11889/j.0253-3219.2023.hjs.46.080016}{https://doi.org/10.11889/j.0253-3219.2023.hjs.46.080016}

\bibitem{Gao2023} Z.-P. Gao and Q.-F. Li, \textit{Studies on several problems in nuclear physics by using machine learning}.
Nucl. Tech. 46, 95 (2023).
\href{https://doi.org/10.11889/j.0253-3219.2023.hjs.46.080009}{https://doi.org/10.11889/j.0253-3219.2023.hjs.46.080009}

\bibitem{Xu_2021} J. Xu, \textit{Constraining isovector nuclear interactions with giant dipole resonance and neutron skin in $^{208}$Pb from a Bayesian approach}. Chin. Phys. Lett. 38, 042101 (2021).
\href{https://dx.doi.org/10.1088/0256-307X/38/4/042101}{https://doi.org/10.1088/0256-307X/38/4/042101}

\bibitem{He2023} W.-B. He,  Q.-F. Li, Y.-G. Ma, Z.-M. Niu, J.-C. Pei, Y.-X. Zhang, \textit{Machine learning in nuclear physics at low and intermediate energies}. Sci. China Phys. Mech. Astron. 66, 282001 (2023).
\href{https://doi.org/10.1007/s11433-023-2116-0}{https://doi.org/10.1007/s11433-023-2116-0}


\bibitem{XU2022137333} J.-Y. Xu, Z.-Z. Li, B.-H. Sun $et~al$., \textit{Constraining equation of state of nuclear matter by charge-changing cross section measurements of mirror nuclei}. Phys. Lett. B 833, 137333
(2022).
\href{https://www.sciencedirect.com/science/article/pii/S0370269322004671}{https://doi.org/10.1016/j.physletb.2022.137333}

\bibitem{ZHAO2023138269}  J.-W. Zhao, B.-H. Sun, I. Tanihata $et~al$., \textit{Isospin-dependence of the charge-changing cross-section shaped by the charged-particle evaporation process}. Phys. Lett. B 847, 138269 (2023).
\href{https://www.sciencedirect.com/science/article/pii/S0370269323006032}{https://doi.org/10.1016/j.physletb.2023.138269}

\bibitem{ghosh2023neutron} T. Ghosh, Sangeeta, G. Saxena, B. K. Agrawal $et~al$., \textit{Neutron Skin Thickness Dependence of Astrophysical $S$-factor}. arXiv:2303.12156 [nucl-th], (2023).
\href{https://arxiv.org/abs/2303.12156}{https://doi.org/10.48550/arXiv.2303.12156}

\bibitem{DONG2019133} J.-M. Dong, X.-L. Shang, W. Zuo $et~al$.,  \textit{An effective Coulomb interaction in nuclear energy density functionals}. Nucl. Phys. A 983, 133 (2019).
\href{https://www.sciencedirect.com/science/article/pii/S0375947419300053}{https://doi.org/10.1016/j.nuclphysa.2019.01.003}

\bibitem{PhysRevC.105.L021304} T. Naito, G. Col\`{o}, H.-Z. Liang $et~al$., \textit{Toward $ab$ $initio$ charge symmetry breaking in nuclear energy density functionals}. Phys. Rev. C 105, L021304 (2022).
\href{https://link.aps.org/doi/10.1103/PhysRevC.105.L021304}{https://doi.org/10.1103/PhysRevC.105.L021304}

\bibitem{PhysRevC.107.064302} T. Naito, G. Col\`{o}, H.-Z. Liang $et~al$.,  \textit{Effects of Coulomb and isospin symmetry breaking interactions on neutron-skin thickness}. Phys. Rev. C 107, 064302 (2023).
\href{https://link.aps.org/doi/10.1103/PhysRevC.107.064302}{https://doi.org/10.1103/PhysRevC.107.064302}

\bibitem{SENG2023137654} C.-Y. Seng and M. Gorchtein, \textit{Electroweak nuclear radii constrain the isospin breaking correction to Vud}. Phys. Lett. B 838, 137654 (2023).
\url{https://www.sciencedirect.com/science/article/pii/S0370269322007882}{https://doi.org/10.1016/j.physletb.2022.137654}

\bibitem{PhysRevC.94.064326} Z. Zhang and L.-W. Chen, \textit{Extended Skyrme interactions for nuclear matter, finite nuclei, and neutron stars}. Phys. Rev. C 94, 064326 (2016).
\href{https://link.aps.org/doi/10.1103/PhysRevC.94.064326}{https://doi.org/10.1103/PhysRevC.94.064326}
\end{thebibliography}
\end{document}